\documentclass[12pt]{article}
\usepackage{amsmath}
\input amssym

\begin{document}

\def\prg#1{\medskip{\bf #1}}     \def\ra{\rightarrow}
\def\lra{\leftrightarrow}        \def\Ra{\Rightarrow}
\def\nin{\noindent}              \def\pd{\partial}
\def\dis{\displaystyle}          \def\inn{\,\rfloor\,}
\def\grl{{GR$_\Lambda$}}         \def\vsm{\vspace{-9pt}}
\def\Lra{{\Leftrightarrow}}
\def\cs{{\scriptstyle\rm CS}}    \def\ads3{{\rm AdS$_3$}}
\def\Leff{\hbox{$\mit\L_{\hspace{.6pt}\rm eff}\,$}}
\def\bull{\raise.25ex\hbox{\vrule height.8ex width.8ex}}
\def\ric{{(Ric)}}                \def\tmgl{\hbox{TMG$_\Lambda$}}
\def\Lie{{\cal L}\hspace{-.7em}\raise.25ex\hbox{--}\hspace{.2em}}
\def\sS{\hspace{2pt}S\hspace{-0.83em}\diagup}
\def\nb{\marginpar{$\bullet$}}   \def\hd{{^\star}}

\def\G{\Gamma}        \def\S{\Sigma}        \def\L{{\mit\Lambda}}
\def\D{\Delta}        \def\Th{\Theta}
\def\a{\alpha}        \def\b{\beta}         \def\g{\gamma}
\def\d{\delta}        \def\m{\mu}           \def\n{\nu}
\def\th{\theta}       \def\k{\kappa}        \def\l{\lambda}
\def\vphi{\varphi}    \def\ve{\varepsilon}  \def\p{\pi}
\def\r{\rho}          \def\Om{\Omega}       \def\om{\omega}
\def\s{\sigma}        \def\t{\tau}          \def\eps{\epsilon}
\def\nab{\nabla}      \def\btz{{\rm BTZ}}   \def\heps{\hat\eps}

\def\tG{{\tilde G}}
\def\cL{{\cal L}}     \def\cM{{\cal M }}   \def\cE{{\cal E}}
\def\cH{{\cal H}}     \def\hcH{\hat{\cH}}
\def\cK{{\cal K}}     \def\hcK{\hat{\cK}}  \def\cT{{\cal T}}
\def\cO{{\cal O}}     \def\hcO{\hat{\cal O}} \def\cV{{\cal V}}
\def\tom{{\tilde\omega}}  \def\cE{{\cal E}}
\def\cR{{\cal R}}    \def\hR{{\hat R}{}}   \def\hL{{\hat\L}}
\def\tb{{\tilde b}}  \def\tA{{\tilde A}}   \def\tv{{\tilde v}}
\def\tT{{\tilde T}}  \def\tR{{\tilde R}}   \def\tcL{{\tilde\cL}}
\def\nn{\nonumber}
\def\be{\begin{equation}}             \def\ee{\end{equation}}
\def\ba#1{\begin{array}{#1}}          \def\ea{\end{array}}
\def\bea{\begin{eqnarray} }           \def\eea{\end{eqnarray} }
\def\beann{\begin{eqnarray*} }        \def\eeann{\end{eqnarray*} }
\def\beal{\begin{eqalign}}            \def\eeal{\end{eqalign}}
\def\lab#1{\label{eq:#1}}             \def\eq#1{(\ref{eq:#1})}
\def\bsubeq{\begin{subequations}}     \def\esubeq{\end{subequations}}
\def\bitem{\begin{itemize}}           \def\eitem{\end{itemize}}
\renewcommand{\theequation}{\thesection.\arabic{equation}}

\title{3D gravity with propagating torsion: the AdS sector}

\author{M. Blagojevi\'c and B. Cvetkovi\'c\footnote{
        Email addresses: {\tt mb@ipb.ac.rs,
                                cbranislav@ipb.ac.rs}} \\
Institute of Physics, P. O. Box 57, 11001 Belgrade, Serbia}
\date{January 20, 2012}
\maketitle

\begin{abstract}
We study the general parity-preserving model of three-dimensional
gravity with propagating torsion, with a focus on its nonlinear
dynamics in the AdS sector. The model is shown to posses the black hole
with torsion as a vacuum solution, provided we adopt certain
restrictions on its coupling constants. The canonical analysis in the
AdS asymptotic region yields the conserved charges of the black hole
and the central charges of the asymptotic conformal symmetry.
\end{abstract}

\section{Introduction}
\setcounter{equation}{0}

In our attempts to properly understand basic aspects of the
gravitational dynamics at both classical and quantum level, we are
naturally led to consider three-dimensional (3D) gravity as a
technically simpler model with the same conceptual features. Following
a traditional approach based on general relativity (GR), 3D gravity has
been studied mainly in the realm of Riemannian geometry, leading to a
number of outstanding results \cite{x1}. In the early 1990s, a new
approach to 3D gravity has been initiated by Mielke and Baekler
\cite{x2}. The approach is based on a modern gauge-field theoretic
conception of gravity characterized by a Riemann--Cartan geometry of
spacetime; it is known as \emph{Poincar\'e gauge theory} (PGT), the
theory in which both the \emph{torsion} and the \emph{curvature} carry
the dynamics of gravity, see \cite{x3,x4,x5,x6}.

The Mielke--Baekler (MB) model is introduced as a \emph{topological} 3D
gravity with torsion, with an idea to explore the influence of geometry
on the dynamics of gravity. Recent investigations along these lines led
to remarkable results: (i) the MB model possesses the black hole
solution, (ii) it can be formulated as a Chern--Simons gauge theory,
(iii) in the AdS sector, asymptotic symmetry is described by two
independent Virasoro algebras with different central charges, (iv) the
black hole entropy is found to depend on torsion, and (v) the geometric
idea of torsion is compatible with supersymmetry; see
\cite{x7,x8,x9,x10,x11}.

Einstein's GR in 3D, with or without a cosmological constant, is also a
topological theory, which has no propagating degrees of freedom. Such a
degenerate situation is not quite a realistic feature of the
gravitational dynamics. Thus, one is naturally motivated to study
gravitational models \emph{with propagating degrees of freedom}. In the
context of Riemannian geometry, there are two well-known models of this
type: topologically massive gravity \cite{x12}, and the
Bergshoeeff--Hohm--Townsend (BHT) massive gravity \cite{x13}. In 3D
gravity with torsion, an extension that includes propagating modes is
even more natural---it corresponds to Lagrangians which are
\emph{quadratic} in the field strengths, as in the standard gauge
approach.

In the present paper, we begin an investigation of the
parity-preserving 3D gravity with propagating torsion. Compared to the
topological MB model \cite{x2}, here we have a Lagrangian with a rather
large number of parameters, and one is faced with the problem of choice
of a set of parameters which defines an acceptable gravitational model.
Some aspect of this problem have been discussed in the literature.
Motivated by the actual importance of massive gravity in high-energy
physics and cosmology, Hernaski et al. \cite{x14} used the spin
projection operators 
to investigate how the existence of propagating torsion can be used to
build up a unitary massive gravity model of the BHT-type, in any
dimension. After that, Helay\"el-Neto et al. \cite{x15} studied the
parity-preserving 3D gravity with propagating torsion combined with the
Chern-Simons term; using the requirements of no ghosts and no tachions,
they found certain restrictions on the parameters. Although these
arguments are commonly accepted in the literature, one should note that
they essentially rely on the weak-field approximation of the theory, in
which inherently nonlinear properties of gravity remain untachable.

Our approach to 3D gravity with propagating torsion is aimed at
studying essential aspects of its nonlinear dynamics. In this paper, we
start by introducing basic elements of the Lagrangian formalism,
whereupon we focus our attention on the AdS sector, examining the
existence of black holes and the nature of asymptotic symmetries. In
the next paper \cite{x16}, we will use the criterion of stability of
the canonical structure under linearization \cite{x17} to find out the
restrictions on parameters that define viable PGT models.

The paper is organized as follows. In section 2, we describe Lagrangian
dynamics of the general parity-preserving PGT in 3D. In section 3, we
give a brief account of the linearized theory, showing that masses of
the propagating modes coincide with those found in \cite{x15}. In
section 4, we find the restrictions on parameters that allow the
existence of the black hole with torsion. In section 5, we apply the
Hamiltonian formalism to construct the canonical gauge generator of the
theory. Then, in section 6, we introduce the AdS asymptotic conditions
and calculate the improved form of the canonical generator, defined by
suitable surface terms. The improved generator is used to find the
conserved charges of the black hole with torsion. Moreover, the
asymptotic symmetry, defined by the canonical algebra of the improved
generators, is described by two independent Virasoro algebras with
central charges, the values of which depend on the quadratic piece of
the Lagrangian.

Our conventions are as follows: the Latin indices $(i,j,k,...)$ refer
to the local Lorentz frame, the Greek indices $(\m,\n,\l,...)$ refer to
the coordinate frame, and both run over 0,1,2; the metric components in
the local Lorentz frame are $\eta_{ij}=(+,-,-)$; totally antisymmetric
tensor $\ve^{ijk}$ is normalized to $\ve^{012}=1$.

\section{General Lagrangian formalism}
\setcounter{equation}{0}

Theory of gravity with torsion can be naturally described as a
Poincar\'e gauge theory, with an underlying Riemann-Cartan (RC)
geometry of spacetime \cite{x3,x4,x5,x6}. Basic gravitational variables
in PGT are the triad field $b^i$ and the Lorentz connection $A^{ij} =
-A^{ji}$ (1-forms), and the corresponding field strengths are
$T^i:=\nabla b^i$ and $R^{ij}:=dA^{ij}+A^i{_k}\wedge A^{kj}$ (2-forms).
The covariant derivative $\nabla=d+\frac{1}{2}A^{ij}\S_{ij}$ (1-form)
acts on a tangent-frame spinor/tensor in accordance with its
spinorial/tensorial structure, reflected in the form of the
representation of the spin matrix $\S_{ij}$.

The antisymmetry of the Lorentz connection $A^{ij}$ implies that the
geometric structure of PGT corresponds to a RC geometry, in which $b^i$
is an orthonormal coframe, $g :=\eta_{ij}b^i\otimes b^j$ is the metric
of spacetime, $A^{ij}$ is the metric-compatible connection defined by
$\nabla g=0$, and $T^i$ and $R^i$ are the torsion and the RC curvature,
respectively.

In local coordinates $x^\m$, we can write $b^i = b^i{_\m}dx^\m$, the
frame dual to $b^i$ reads $h_i = h_i{^\m}\pd_\m$, and we have $h_i\inn
b^j = h_i{^\m}b^j{_\m} = \d^i_j$, where $\inn$ is the interior product.

\subsection{Lagrangian and the field equations}

General dynamics of 3D gravity with propagating torsion is defined by
the Lagrangian 3-form
\bsubeq\lab{2.1}
\be
L=L_G(b^i,T^i,R^{ij})+L_M(b^i,\psi,\nabla\psi)             \lab{2.1a}
\ee
where $L_M$ denotes matter contribution, and the gravitational piece
$L_G$ is at most quadratic in torsion and curvature. Assuming that
$L_G$ preserves parity, we have
\bea
L_G&=&-a\ve_{ijk}b^i\wedge R^{jk}
  -\frac{1}{3}\L_0\ve_{ijk}b^i\wedge b^j\wedge b^k
  +L_{T^2}+L_{R^2}\, ,                                     \nn\\
L_{T^2}&=&T^i\wedge
  \hd\left(a_1{}^{(1)}T_i+a_2{}^{(2)}T_i+a_3{}^{(3)}T_i\right)\,,\nn\\
L_{R^2}&=&\frac{1}{2}R^{ij}\wedge\hd\left(b_4{}^{(4)}R_{ij}
  +b_5{}^{(5)}R_{ij}+b_6{}^{(6)}R_{ij}\right)\, ,          \lab{2.1b}
\eea
\esubeq
where $^{(a)}T_i$ and $^{(a)}R_{ij}$ are irreducible components of the
torsion and the RC curvature, see Appendix A. In what follows, we will
omit the wedge product sign $\wedge$ for simplicity.

Let us now introduce the covariant gravitational momenta (1-forms)
\be
H_i:=\frac{\pd L_G}{\pd T^i}\, ,\qquad
H_{ij}:=\frac{\pd L_G}{\pd R^{ij}}\, .
\ee
In addition to that, we define the dynamical energy-momentum and spin
currents (2-forms) for the gravitational field:
\be
t_i:=\frac{\pd L_G}{\pd b^i}\, ,\qquad
s_{ij}:=\frac{\pd L_G}{\pd A^{ij}}\, ,
\ee
as well as the corresponding matter currents (2-forms):
\be
\t_i:=\frac{\pd L_M}{\pd b^i}\, ,\qquad
\s_{ij}:=\frac{\pd L_M}{\pd A^{ij}}
         =\S_{ij}\psi\,\frac{\pd L_M}{\pd\nabla\psi}\,.
\ee
Then, using the relations \cite{x6}
\be
\d T^i=\nabla\d b^i+\d A^{ij}\wedge b_j\, ,\qquad
\d R^{ij}=\nabla\d A^{ij}\, ,                            \nn
\ee
one finds that the variation of the Lagrangian \eq{2.1a} with respect
to $b^i$ and $A^{ij}$ produces the following gravitational field
equations:
\bsubeq\lab{2.5}
\bea
&&\nabla H_i+t_i=-\t_i\, ,                                 \\
&&\nabla H_{ij}+s_{ij}=-\s_{ij}\, .
\eea
\esubeq
Explicit calculation based on the gravitational Lagrangian \eq{2.1b}
yields
\bea
&&H_i=2\hd\left(
     a_1{}^{(1)}T_i+a_2{}^{(2)}T_i+a_3{}^{(3)}T_i\right)\,,\nn\\
&&H_{ij}=-2a\ve_{ijk}b^k+H'_{ij}\, ,                       \nn\\
&&H'_{ij}:=2\hd\left(b_4{}^{(4)}R_{ij}
           +b_5{}^{(5)}R_{ij}+b_6{}^{(6)}R_{ij}\right)\, ,
\eea
and
\bea
t_i&=&e_i\inn L_G-(e_i\inn T^m)\wedge H_m
      -\frac{1}{2}(e_i\inn R^{mn})\wedge H_{mn}\, ,        \nn\\
s_{ij}&=&-\left(b_i\wedge H_j-b_j\wedge H_i\right)\, .
\eea
The second field equation can be now rewritten in an equivalent form
as:
$$
-2a\ve_{ijk}T^k+\nabla H'_{ij}+s_{ij}=-\s_{ij}\, .   \eqno(2.5b')
$$

Using the above expressions for the gravitational field momenta, the
gravitational Lagrangian can be written in a more compact form as:
\be
L=\frac{1}{2}T^iH_i+\frac{1}{2}R^{ij}(-2a\ve_{ijk}b^k)
  +\frac{1}{4}R^{ij}H'_{ij}
  -\frac{1}{3}\L_0\ve_{ijk}b^ib^jb^k\, .                   \lab{2.8}
\ee

Bianchi identities for PGT read:
\be
\nabla T^i=R^i{_j}b^j\, ,\qquad \nabla R^{ij}=0\, .
\ee
For the form of Noether identities, we refer the reader to Ref.
\cite{x6}.

In 3D gravity, the Weyl curvature vanishes:
\be
W_{ij}=R_{ij}-(b_i\hR_j-b_j\hR_i)+\frac{1}{2}Rb_ib_j=0\, . \lab{2.10}
\ee
As a consequence, the RC curvature $R_{ij}$ can be expressed in terms
of the Ricci 1-form $\hR_i=\hR_{ij}b^j$ (or more compactly, in terms of
the Schouten 1-form $L_i=\hR_i-\frac{1}{4}Rb_i$). This property can be
easily accommodated in the general Lagrangian formalism.

\subsection{The tensor formalism}

For later convenience and an easier comparison with the results in the
literature, we present here the tensor form of Lagrangian and the field
equations, see Appendix A and \cite{x3}.

Using the notation $\tcL=b\cL$, the gravitational Lagrangian reads
\bsubeq
\be
\cL_G=-aR-2\L_0+\cL_{T^2}+\cL_{R^2}\, ,
\ee
where
\bea
\cL_{T^2}&:=&\frac{1}{2}T^{ijk}\left(a_1{}^{(1)}T_{ijk}
           +a_2{}^{(2)}T_{ijk}+a_3{}^{(3)}T_{ijk}\right)   \nn\\
\cL_{R^2}&:=&\frac{1}{4}R^{ijkl}\left( b_4{}^{(4)}R_{ijkl}
        +b_5{}^{(5)}R_{ijkl}+b_6{}^{(6)}R_{ijkl}\right)\,.
\eea
\esubeq
The expansions $H_i=\frac{1}{2}\cH_{ijk}\hd(b^jb^k)$ and
$H_i=\frac{1}{2}\cH_{ijk}\hd(b^jb^k)$ define the components of the
gravitational momenta as:
\bea
\cH_{ijk}&=&2\left(a_1{}^{(1)}T_{ijk}
       +a_2{}^{(2)}T_{ijk}+a_3{}^{(3)}T_{ijk}\right)\, ,   \nn\\
\cH_{ijkl}&=&
  =-2a(\eta_{ik}\eta_{jl}-\eta_{jk}\eta_{il})+\cH'_{ijkl}\,,\nn\\
\cH'_{ijkl}&=&2\left(b_4{}^{(4)}R_{ijkl}
     +b_5{}^{(5)}R_{ijkl}+b_6{}^{(6)}R_{ijkl}\right)\, .   \nn
\eea
Note that $\cH_{ijk}={\pd\cL_G}/{\pd T^{ijk}},
\cH_{ijkl}={\pd\cL_G}/{\pd R^{ijkl}}$. Then, the field equations read
\bsubeq
\bea
&&\nabla_\m H_i{}^{\m\n}-t_i{^\n}=\t_i{^\n}\, ,          \\
&&\nabla_\m H_{ij}{}^{\m\n}-s_{ij}{}^\n=\s_{ij}{^\n}\,,
\eea
\esubeq
where  $H_{ijk}=b\cH_{ijk}$, $H_{ijkl}=b\cH_{ijkl}$, and
\bea
&&t_i{^\n}=h_i{^\n}\tcL_G
  -H_{mn}{^\n}T^{mn}{_i}-\frac{1}{2}H_{mnr}{^\n}R^{mnr}{_i}\,,\nn\\
&&s_{ij}{^\n}=-\left(H_{ij}{^\n}-H_{ji}{^\n}\right)\, . \nn
\eea
Isolating the contribution of the linear curvature term in $H_{ijkl}$,
the field equations read:
\bsubeq\lab{2.13}
\bea
&&\nabla_\m H_i{}^{\m\n}
  -h_i{^\n}\tcL_G+H_{mn}{^\n}T^{mn}{_i}-2abR^\n{_i}
  +\frac{1}{2}H'_{mnk}{^\n}R^{mnk}{_i}=\t_i{^\n}\,,        \\
&&-a\ve^{\m\n\r}_{ijk}T^k{}_{\m\r}+\nabla_\m H'_{ij}{}^{\m\n}
  +2H_{[ij]}{^\n}=\s_{ij}{^\n}\, .
\eea
\esubeq

We find it useful to give here an equivalent description of the $T^2$
Lagrangian:
\bea
&&\cL_{T^2}=\frac{1}{4}T^{ijk}\cH_{ijk}=T^{ijk}\left(
   \a_1 T_{ijk}+\a_2 T_{kji}+\a_3\eta_{ij}v_k\right)\, ,   \\
&&\a_1=\frac{1}{6}(2a_1+a_3)\, ,\quad\a_2=\frac{1}{3}(a_1-a_3)\, ,
  \quad \a_3=\frac{1}{2}(a_2-a_1)\, .                      \nn
\eea
Moreover, the vanishing of the Weyl curvature implies that both
$\cH'_{ijkl}$ and $\cL_{R^2}=\frac{1}{8}R^{ijkl}\cH'_{ijkl}$ can be
expressed in terms of the Ricci tensor $\hR_{ij}$:
\bea
&&\cH'_{ijkl}=
   2(\eta_{ik}\g_{jl}-\eta_{jk}\g_{il})-(k\lra l)\, ,      \nn\\
&&\cL_{R^2}=\hR^{ij}\left(\b_1\hR_{ij}+\b_2\hR_{ji}
  +\b_3\eta_{ij}R\right)=:\hR^{ij}\g_{ij}\, ,              \\
&&\b_1=\frac{1}{2}(b_4+b_5)\, ,\quad \b_2=\frac{1}{2}(b_4-b_5)\, ,
     \quad \b_3=\frac{1}{12}(b_6-4b_4)\, .                 \nn
\eea

\subsection{Lie dual forms}

Let us note that in 3D, for any antisymmetric form $X^{ij}=-X^{ji}$,
one can define its \emph{Lie dual} form $X_k$ by
$X^{ij}=-\ve^{ijk}X_k$. It is often convenient to replace
($A^{ij},R^{ij}$) by the corresponding Lie duals ($\om^i,R^i$), so that
$$
T^i=db^i+\ve^i{}_{jk}\om^j b^k\, ,\qquad
R^i=d\om^i+\frac{1}{2}\ve^i{}_{jk}\om^j\om^k\, .
$$
We will switch to this Lie dual notation in section 4.

\section{Particle spectrum on Minkowski background}
\setcounter{equation}{0}

The weak-field approximation of 3D gravity with propagating torsion
around the Minkowski background $M_3$ yields an approximate picture of
the gauge structure and dynamical content of the theory. To what extent
this picture reflects essential features of the full, nonlinear theory
is an issue that will be examined in \cite{x16}.

The particle spectrum of 3D gravity with propagating torsion has been
studied by Helay\"{e}l-Neto et. al. \cite{x15} in a model based on the
Lagrangian \eq{2.1} plus a parity-violating Chern-Simons term. Using an
extended basis of the spin projection operators, they were able to
examine the conditions for the theory to have well-behaved propagating
modes (no ghosts and no tachions). Here, we shall re-derive these
results for the theory \eq{2.1}, using an approach based on the
covariant field equations, see \cite{x19}.

The only propagating modes of the theory \eq{2.1} are those associated
to the Lorentz connection $A^{ij}{_\m}$. In the weak field
approximation, $A^{ij}{_\m}$ has 9 independent modes. For massive
modes, the spin content can be determined by looking at the
corresponding irreducible representations of the little group $SO(2)$.
Since $SO(2)$ is Abelian, all its representations are
\emph{one-dimensional}. By subtracting 3 gauge degrees of freedom,
corresponding to 3 local Lorentz rotations, one finds that at most 6
degrees of freedom can be physical.

In two spatial dimensions, \emph{parity} is defined as the inversion of
one axis, the inversion of both would be a rotation \cite{x20}; thus,
for instance, we can define it as the inversion of $y$ axis. Since
parity is a symmetry of the Lagrangian \eq{2.1}, the structure of the
corresponding irreducible representations is changed: they contain
\emph{two states} with the same value of spin $J$, which transform into
each other under $P$.

The irreducible representations of $SO(2)\times P$ provide a foundation
for understanding the particle content of any Lorentz-covariant field
theory in 3D. However, it is usually simpler to start with
finite-dimensional representations of the full Lorentz group $SO(1,2)$.
Since these covariant fields are not, in general, irreducible,  they
are combined with the field equations and subsidiary conditions so as
to remove the unphysical degrees of freedom \cite{x21}.

For $\L_0=0$, the Minkowski configuration $(b^i{_\m},
A^{ij}{_\m})=(\d^i_\m,0)$ is a solution of the field equations
\eq{2.13} in vacuum. The weak-field approximation around $M_3$ takes
the form
\bsubeq
\be
b^i{_\m}=\d^i_\m+\tb^i{_\m}\,,\qquad A^{ij}{_\m}=\tA^{ij}{}_\m\,,
\ee
where the tilde sign denotes small field excitations. Then:
\bea
&&T^i{}_{\m\n}=\tilde T^i{}_{\m\n}+\cO_2\,,\qquad
  \tilde T^i{}_{\m\n}=
  \pd_\m\tb^i{_\n}-\pd_\n\tb^i{_\m}+2\tA^i{}_{[\n\m]}\,,   \nn\\
&& R^{ij}{}_{\m\n}=\tR^{ij}{}_{\m\n}+\cO_2\,,\qquad
  \tR^{ij}{}_{\m\n}=\pd_\m\tA^{ij}{}_\n-\pd_\n\tA^{ij}{}_\m\,.
\eea
\esubeq
The linearized field equations take the form:
\bsubeq
\bea
\pd_\m\tilde H_i{}^{\m\n}-2a\tilde G^\n{_i}
 &=&\tilde\t_i{^\n}\, ,                                    \lab{3.2a}\\
-a\ve^{\m\n\r}_{ijk}\tilde T^k{}_{\m\r}
  +\pd_\m\tilde H'_{ij}+2\tilde H_{[ij]}{^\n}
  &=&\tilde\s_{ij}{^\n}\, .                                \lab{3.2b}
\eea
\esubeq

Being basically interested in spins and masses of the tordion modes, we
restrict our attention to the vacuum field equations. Moreover, in what
follows we omit tilde for simplicity. Using \eq{3.2a} we can express
the Ricci curvature in terms of the first derivatives of the torsion
tensor, whereupon \eq{3.2b} can be transformed into a set of equations
containing the torsion tensor and its second derivatives. A suitable
``diagonalization" of these equations leads to several Klein--Gordon
equations for the torsion modes. Here, we consider the case when these
modes are \emph{massive}.

(a) The field $a=\frac{1}{6}\ve_{ijk}T^{ijk}$ satisfies the
Klein--Gordon equation $(\square+m_{0^-}^2)a=0$ with
\be
m_{0^-}^2=\frac{3(a-a_1)(a+2a_3)}{(a_1+2a_3)b_5}\,.
\ee
Thus, $a$ is a massive pseudoscalar state, $J^P=0^-$.

(b) Similarly, the field $\s=\pd^i v_i$ is a massive scalar with

\be
m_{0^+}^2=\frac{3a(a+a_2)}{a_2(b_4+2b_6)}\, ,
\ee
and $J^P=0^+$.

(c) For $\bar v^i:=v^i+\dis\frac{1}{m_{0^+}^2}\pd^i\s$ (with $\pd^k
\bar v_k=0$), we find:
\be
m_{1}^2=\frac{4(a-a_1)(a+a_2)}{(a_1+a_2)(b_4+b_5)}\,.
\ee
The transverse field $\bar v_i$ describes two massive states with spin
$J=1$; these two states transform into each other under parity (like
the helicity states in 4D).

(d) The field
$$
\chi_{ij}=\pd^k  t_{k(ij)}
 +\frac{a_1(a+a_2)^2}{2(a-a_1)\left[a(a_2-a_1)-2a_1a_2\right]}
  \pd_{(i}\bar v_{j)}\,,
$$
is symmetric, traceless and divergenceless, and it satisfies the
Klein--Gordon equation with
\be
m_{2}^2=-\frac{a(a-a_1)}{a_1b_4}\,.
\ee
Hence, $\chi_{ij}$ describes two massive degrees of freedom with spin
$J=2$. Again, these two states form a parity invariant multiplet.

Using the field equations and the identities
$\ve^{ijk}t_{ijk}=t^i{}_{ik}=0$, one can show that $t_{ijk}$ and its
first and second derivatives can be expressed in terms of ($a$, $\s$,
$\bar v_i$, $\chi_{ij}$) and their first and second derivatives. Thus,
the spectrum of excitations around $M_3$ consists of \emph{6
independent torsion modes}: two spin-$0^\mp$ states ($a,\s$), two
spin-1 states $\bar v_i$, and two spin-2 states $\chi_{ij}$, in
agreement with the little group analysis. Our results coincide with
those obtained in \cite{x15}, in the limit when the Chern--Simons
coupling constant vanishes.

\section{The black hole with torsion}
\setcounter{equation}{0}

As a first step in our study of the nonlinear dynamics of 3D gravity
with propagating torsion, we wish to examine whether the PGT model
\eq{2.1} admits the AdS black hole solution.

In the MB model of 3D gravity with torsion \cite{x2}, there exists an
interesting vacuum solution, the black hole with torsion
\cite{x7,x8,x9}. In the Schwarzschild-like coordinates
$x^\m=(t,r,\vphi)$, this solution is defined by the pair $(b^i,\om^i)$,
where (a) the triad field has the form
\bsubeq\lab{4.1}
\be
b^0=Ndt\, ,\qquad b^1=N^{-1}dr\, ,\qquad b^2=r(d\vphi+N_\vphi dt)\, ,
\ee
where $N$ and $N_\vphi$ are the lapse and the shift functions of the
Ba\~nados-Teitelboim-Zanelli (BTZ) metric \cite{x22} ($m$ and $J$ are
the integration constants):
$$
N^2=-8mG+\frac{r^2}{\ell^2}+\frac{16G^2J^2}{r^2}\, ,\qquad
N_\vphi=\frac{4GJ}{r^2}\, ,
$$
and (b) the connection $\om^i$ can be found as the solution of the MB
vacuum field equations:
\be
\om^i=\tilde\om^i+\frac{p}{2}b^i\, ,
\ee
\esubeq
where $\tilde\om^i$ is the Riemannian connection, and $p$ is a
parameter that measures torsion. For the solution \eq{4.1}, the field
strengths have the following form:
\be
2T_i=p \ve_{ijk}b^jb^k\,, \qquad 2R_i=q\ve_{ijk}b^jb^k\, , \lab{4.2}
\ee
where $p$ and $q$ are parameters, and we assume that the effective
cosmological constant is negative \cite{x8,x9}:
$$
\Leff=q-\frac{1}{4}p^2=-\frac{1}{\ell^2}<0\, .
$$

In the context of RC geometry, the form-invariance of a given field
configuration is defined by the requirements $\d_0 b^i{_\m}=
\d_0\om^i{_\m}=0$, which differ from the Killing equation in GR ($\d_0$
is the PGT analogue of the Lie derivative). Symmetry properties of the
black hole \eq{4.1} are expressed by its form-invariance under the
action of two Killing vectors, $\pd/\pd t$ and $\pd/\pd\vphi$. In a RC
spacetime with $\Leff<0$, there exists another, maximally symmetric
solution, known as the \emph{AdS solution} (AdS$_3$), which can be
formally obtained from \eq{4.1} by the replacement $J=0$, $2m=-1$. The
form-invariance of AdS$_3$ is described by the six-dimensional AdS
group $SO(2,2)$. The black hole and AdS$_3$ are locally isometric, but
globally distinct solutions \cite{x9}.

By combining \eq{4.2} with the field equations \eq{2.13} in vacuum, we
can obtain certain restrictions on $p$ and $q$, under which the
BTZ-like black hole, as well as AdS$_3$, is an exact solution of the
theory. To find these these restrictions, we use \eq{4.2} to obtain
\bea
&&H_i=2pa_3b_i\, ,                                         \nn\\
&&t_i=\left(aq-\L_0-\frac{1}{2}p^2a_3
            -\frac{1}{2}q^2b_6\right)\ve_{ijk}b^jb^k\, ,   \nn\\
&&H_{ij}=-2(a+qb_6)\ve_{ijk}b^k\, ,                        \nn\\
&&s_{ij}=-4pa_3b_ib_j\, ,                                  \nn
\eea
whereupon the field equations \eq{2.13} in vacuum lead to:
\bea
&&aq-\L_0+\frac{1}{2}p^2a_3-\frac{1}{2}q^2b_6=0\,,         \nn\\
&&p(a+qb_6+2a_3)=0\, .
\eea
These conditions guarantee that the black hole with torsion \eq{4.1} is
a solution of the PGT model \eq{2.1}. The second equation naturally
leads to the following two cases:
\bitem
\item[a)] $p=0$ $\quad\Ra\quad$\\
For $b_6\ne 0$, we have
$$
qb_6=a\pm\sqrt{a^2-2b_6\L_0}\,.
$$
If, additionally, $a^2-2b_6\L_0=0$, the value of $qb_6$ is unique:
$qb_6=a$.\\
For $b_6=0$, the value of $q$ is $q=\L_0/a$.
\item[b)] $a+qb_6+2a_3=0$ $\quad\Ra\quad$
$$
\frac{1}{2}a_3p^2=\L_0+\frac{1}{2}q(qb_6-2a)
                 =\L_0+\frac{1}{2b_6}(2a_3+a)(2a_3+3a)\,.
$$
For $a_3=0$, $p$ remains undetermined, which is physically not
acceptable.
\eitem

\section{Canonical gauge generator}
\setcounter{equation}{0}

As an important step in our examination of the asymptotic structure of
spacetime, we are going to construct the canonical gauge generator,
which is our basic tool for studying asymptotic symmetries and
conserved charges of 3D gravity with propagating torsion.

\subsection{First-order formalulation}

Following Nester's ideas \cite{x23}, based on the analogy with the
first-order formulation of electrodynamics, $L_{ED}=dA\wedge\hd
F+\frac{1}{2}F\wedge \hd F$, we introduce the \emph{first-order
formulation}\footnote{The term ``first-order formulation" used here
differs from the common usage in gravity, where it refers to the case
in which $b^i$ and $\om^i$ are independent Lagrangian variables (as in
the standard form of PGT).} of the gravitational theory \eq{2.1} by
\bsubeq\lab{5.1}
\be
L=T^i\t_i+R^i\r_i-V(b^i,\t_i,\r'_i)
  -\frac{1}{3}\L_0\ve_{ijk}b^ib^jb^k+L_M\, ,               \lab{5.1a}
\ee
where $\t_i$ and $\r_i:=2ab_i+\r'_i$ are the covariant field momenta
(1-forms), conjugate to $b^i$ and $\om^i$. Here, not only $b^i$ and
$\om^i$, but also $\t_i$ and $\r_i$ are \emph{independent} dynamical
variables. The piece $2ab_i$ in $\r_i$ is a correction stemming from
the term linear in curvature. The potential $V$ is quadratic in $\t$
and $\r'$, and its form is chosen so as to ensure the on-shell
equivalence of the new formulation \eq{5.1a} with \eq{2.1}. Indeed, the
variation with respect to $\t_i$ and $\r_i$ produces the field
equations $\t_i=H_i$ and $\r'_{ij}:=-\ve_{ij}{^k}\r'_k=H'_{ij}$
(Appendix B). The first-order formulation leads to a particularly
simple construction of the gauge generator \cite{x24}.

Our approach is based on the canonical structure of the theory, which
we investigate in the standard tensor calculus. In local coordinates
$x^\m$, the first-order Lagrangian density, corresponding to \eq{5.1a},
has the form:
\be
\tcL=\frac{1}{2}\ve^{\m\n\r}\left(T^i{}_{\m\n}\t_{i\r}
  +R^i{}_{\m\n}\r_{i\r}\right)-\cV(b,\t,\r')+\tcL_M\, ,    \lab{5.1b}
\ee
\esubeq
where $\cV$ stands for the sum of $V$ and the cosmological term. We are
concerned here with finite gravitational sources, characterized by
matter fields which decrease sufficiently fast at large distances, so
that their contribution to surface integrals vanishes. Thus, the
asymptotic dynamics can be studied by ignoring matter fields.

For reference, we display here the vacuum field equations obtained by
varying $\tcL$ with respect to $b^i{_\m},\om^i{_\m},\t^i{_\m},$ and
$\r^i{_\m}$:
\bsubeq\lab{5.2}
\bea
\ve^{\m\n\r}\nabla_\n\t_{i\r}-\frac{\pd\cV}{\pd b^i{_\m}}=0\,,\quad
 &&\ve^{\m\n\r}\left(\nab_\n\r_{i\r}
        +\ve_{ijk}b^j{}_\n \t^k{}_\r\right)=0 \, ,         \lab{5.2a}\\
\frac{1}{2}\ve^{\m\n\r}T_{i\n\r}-\frac{\pd\cV}{\pd\t^i{_\m}}=0\,,\quad
 &&\frac{1}{2}\ve^{\m\n\r}R_{i\n\r}-\frac{\pd\cV}{\pd\r^i{_\m}}=0\,.\lab{5.2b}
\eea
\esubeq
As we mentioned above, equations \eq{5.2b} determine the values of
$\t_{i\m}$ and $\r'_{i\m}$.

\subsection{Hamiltonian and constraints}

\prg{Primary constraints.} By introducing the momenta
$\pi_A:=(\pi_i{^\m},\Pi_i{^\m},p_i{^\m},P_i{^\m})$, conjugate to the
respective Lagrangian variables $\vphi^A:=(b^i{_\m}, \om^i{_\m},
\t^i{_\m}, \r^i{_\m})$, we obtain the following primary constraints:
\bea
&&\phi_i{^0}:=\pi_i{^0}\approx 0\, ,\qquad\,\,
  \phi_i{^\a}:=\pi_i{^\a}-\ve^{0\a\b}\t_{i\b}\approx 0\, , \nn\\
&&\Phi_i{^0}:=\Pi_i{^0}\approx 0\, ,\qquad
  \Phi_i{^\a}:=\Pi_i{^\a}-\ve^{0\a\b}\r_{i\b}\approx 0\,.   \nn\\
&&p_i{^\m}\approx 0\, ,\hspace{61pt} P_i{^\m}\approx 0\, . \lab{5.3}
\eea
The canonical Hamiltonian is conveniently represented as
\be
\cH_c= b^i{}_0\cH_i+\om^i{}_0\cK_i+\t^i{_0}\cT_i+\r^i{_0}\cR_i
         +\cV+\pd_\a D^\a\, ,               \nn\\
\ee
where
\bea
&&\cH_i=-\ve^{0\a\b}\nabla_\a\t_{i\b}\,,\nn\\
&&\cK_i=-\ve^{0\a\b}\left(\nab_\a\r_{i\b}
        +\ve_{ijk}b^j{}_\a \t^k{}_\b\right) \, ,           \nn\\
&&\cT_i=-\frac{1}{2}\ve^{0\a\b}T_{i\a\b}\,,                \nn\\
&&\cR_i=-\frac 12\ve^{0\a\b}R_{i\a\b}\, ,                  \nn
\eea
and $D^\a=\ve^{0\a\b}\left( \om^i{}_0\r_{i\b}+b^i{}_0 \t_{i\b}\right)$.

\prg{Secondary constraints.} Going over to the total Hamiltonian,
\be
\cH_T=\cH_c +u^i{}_\m\phi_i{}^\m+v^i{}_\m\Phi_i{}^\m
            +w^i{}_\m p_i{^\m}+z^i{_\m}P_i{^\m}\, ,        \nn
\ee
where $(u,v,w,z)$ are canonical multipliers, we find that the
consistency conditions of the primary constraints $\pi_i{}^0$,
$\Pi_i{}^0$, $p_i{}^0$ and $P_i{^0}$ yield the secondary constraints:
\bea
&&\hcH_i:=\cH_i+\frac{\pd\cV}{\pd b^i{_0}}\approx 0\, ,    \nn\\
&&\cK_i\approx 0\, ,                                       \nn\\
&&\hat\cT_i:=\cT_i+\frac{\pd \cV}{\pd \t^i{_0}}\approx 0\,,\nn\\
&&\hat \cR_i:=\cR_i+\frac{\pd \cV}{\pd \r^i{_0}}\approx 0\,.\lab{5.4}                                      \lab{A.3}
\eea
They correspond to the $\m=0$ components of the vacuum field equations
\eq{5.2}.

The consistency of the remaining primary constraints
$X_A:=(\phi_i{}^\a, \Phi_i{}^\a, p_i{}^\a, P_i{}^\a)$ leads to the
determination of the multipliers
$(u^i{}_\a,v^i{}_\a,w^i{}_\a,z^i{_\a})$ . However, we find it more
convenient to continue our analysis in a \emph{reduced phase space}.
Using the second class constraints $X_A$, we can eliminate the momenta
$(\pi_i{^\a},\Pi_i{^\a},p_i{^\a},P_i{^\a})$ and construct the reduced
phase space $R$, in which the basic nontrivial Dirac brackets (DBs)
read (Appendix C):
\bea
&&\{b^i{_\a},\t^j{_\b}\}^*=\eta^{ij}\ve_{0\a\b}\d\, ,  \qquad
\{\om^i{_\a},\r^j{_\b}\}^*=\eta^{ij}\ve_{0\a\b}\d \,.\lab{5.5}
\eea
The remaining DBs are the same as the corresponding Poisson brackets
(PBs).

In $R$, the total Hamiltonian is simplified:
\bsubeq
\be
\cH_T=\cH_c +u^i{}_0\phi_i{}^0+v^i{}_0\Phi_i{}^0
            +w^i{}_0 p_i{^0}+z^i{_0}P_i{^0}\, ,
\ee
where $\cH_c$ can be represented in a ``more canonical" form as
\be
\cH_c=b^i{_0}\hat\cH_i+\om^i{_0}\cK_i
      +\t^i{_0}\hat\cT_i+\r^i{_0}\hat\cR_i.
\ee
\esubeq
Note that $\cH_c$ is not linear in
$(b^i{_0},\om^i{_0},\t^i{_0},\r^i{_0})$, as the formula might suggest.

\prg{Further consistency conditions.} In the nest step, we would have
to examine the consistency conditions of the secondary constraints
\eq{5.4}. This may lead to an additional set of tertiary constraints,
and so on. However, it turns out that such an analysis is in fact not
necessary, since we are able to complete our main goal---the
construction of the gauge generator---in a simpler way.

\subsection{Gauge generator}

When the canonical structure of the theory is completely known, the
canonical gauge generator can be constructed using the well-known
Castellani's algorithm \cite{x25}. However, it is sometimes simpler to
use an alternative method, based on the following criterion \cite{x4}:
\bitem
\item[--] a phase-space functional $G$ is the canonical gauge generator
if it produces the correct gauge transformations of all the phase-space
variables.
\eitem

To be able to use this property, we note that in an earlier paper
\cite{x26}, we derived the form of the gauge generator $G$ in the BHT
massive gravity \cite{x13}, a Riemannian theory with quadratic
curvature terms. Looking at the gauge generator (5.2) in \cite{x26},
where $\l_{i\m}$ plays the role of $\t_{i\m}$, and $f_{i\m}$ coincides
with $\r'_{i\m}$, we are naturally led to the conjecture that the gauge
generator of the present theory has the following form:
\bea
G&=&-G_1-G_2\, ,                                           \nn\\
G_1&=&\dot\xi^\m\left(b^i{}_\m\pi_i{}^0
       +\om^i{}_\m\Pi_i{}^0+\t^i{_\m}p_i{}^0
       +\r^i{_\m}P_i{^0}\right)                            \nn\\
&&+\xi^\m\left[b^i{}_\m\hcH_i+\om^i{}_\m\cK_i
               +\t^i{_\m}\hat\cT_i+\r^i{_\m}\hat\cR_i\right.\nn\\
&&\qquad+\left.(\pd_\m b^i{_0})\pi_i{}^0+(\pd_\m\om^i{}_0)\Pi_i{}^0
  +(\pd_\m\t^i{_0})p_i{^0}+(\pd_\m\r^i{_0})P_i{^0}\right]\,,\nn\\
G_2&=&\dot{\th^i}\Pi_i{}^0+\th^i\left[\cK_i
  -\ve_{ijk}\left(b^j{}_0\pi^{k0}+\om^j{}_0\Pi
  +\t^j{}_0p^{k0}+\r^j{}_0P^{k0}\right)\right]\, .        \lab{5.7}
\eea

The proof of this conjecture is presented in Appendix D. Indeed,
looking at the action of $G$ on the phase-space variables $\vphi$ in
$R$, defined by $\d_0\vphi=\{\vphi,G\}^*$, one finds that these gauge
transformations coincide with the Poincar\'e gauge transformations
$\d_{\rm P}\vphi$ on shell.

\section{AdS asymptotic structure}
\setcounter{equation}{0}

Asymptotic conditions imposed on dynamical variables determine the form
of asymptotic symmetries and the corresponding conservation laws. As
before, we assume that $\Leff<0$.

\subsection{Asymptotic conditions}

AdS asymptotic behavior is defined by the following set of requirements
\cite{x27,x9}:
\bitem
\item[(a)] asymptotic configurations should include the black hole
geometries;\vsm
\item[(b)] they should be invariant under the action of the AdS group
$SO(2,2)$;\vsm
\item[(c)] asymptotic symmetries should have well-defined canonical
generators.
\eitem
The requirements (a) and (b) lead to the following asymptotic form of
the triad field and the connection \cite{x9}:
\bsubeq\lab{6.1}
\be
b^i{_\m}=\left( \ba{ccc}
       \dis\frac{r}{\ell}+\cO_1   & O_4  & O_1  \\
       \cO_2 & \dis\frac{\ell}{r}+\cO_3  & O_2  \\
       \cO_1 & \cO_4                     & r+\cO_1
                 \ea
          \right)=:\tilde b^i{_\m}\, ,
\ee
\be
\om^i{_\m}\sim\tom^i{_\m}+\frac{p}{2}\tilde b^i{_\m}\, ,\qquad
\tom^i{_\m}=\left( \ba{ccc}
       \cO_1  & O_2    & -\dis\frac{r}{\ell}+\cO_1  \\
       \cO_2  & \cO_3  & O_2                        \\
       -\dis\frac{r}{\ell^2}+\cO_1 & \cO_2   & \cO_1
                         \ea
          \right)\, ,
\ee
where $\tom^i{_\m}$ is the Riemannian piece of the connection. Then,
using the fact that the field equations \eq{5.2b} imply
$\t_i=H_i$ and $-\ve_{ij}{^k}\r_k=H_{ij}$, we find the AdS asymptotic
behavior of the covariant field momenta:
\be
\t^i{_\mu}\sim 2pa_3\tilde b^i{_\m}\,,\qquad
  \r^i{_\m}\sim 2(a+ qb_6)\tilde b^i{_\m}\, .
\ee
\esubeq

A closer inspection of this result shows a remarkable similarity to the
related structure in the MB model with \emph{vanishing} Chern--Simons
coupling constant ($\a_3=0$), denoted as MB$'$. Indeed, in MB$'$ we have
$\t_i=\a_4 b^i$ and $\r_i=2ab^i$, as follows from equations (2.2) in
\cite{x24}. Thus, although MB$'$ and 3D gravity with propagating
torsion \eq{2.1} have rather different structures in the bulk, their
AdS asymptotic behaviors are related by a simple correspondence:
\be
\a_4\lra 2pa_3\,,\qquad a\lra a+qb_6\, .                   \lab{6.2}
\ee
Clearly, these relations shed a new light on the AdS asymptotic
structure of 3D gravity with propagating torsion.

\subsection{Conserved charges}

The canonical generator, which acts on dynamical variables via the DB
operation, has to be a differentiable phase-space functional. For a
given set of asymptotic conditions, this property is ensured by adding
suitable surface terms to $G$ \cite{x28}.

By adopting the asymptotic conditions \eq{6.1}, the improved canonical
generator \eq{5.7} is found to have the form
\bsubeq\lab{6.3}
\be
\tG=G+\G\,,\quad
\G=-\int_0^{2\pi}d\vphi\left(\xi^0\cE^1+\xi^2\cM^1\right)\,,\lab{6.3a}
\ee
where
\bea
\cE^1&:=&2\left[(a+qb_6)\om^0{_2}
  +\left(2pa_3+\frac{a+qb_6}{2}p\right)b^0{_2}
  +\frac{a+qb_6}\ell b^2{_2}\right]b^0{_0}\,,             \nn\\
\cM^1&:=&-2\left[(a+qb_6)\om^2{_2}
  +\left(2pa_3+\frac{a+qb_6}{2}p\right)b^2{_2}
  + \frac{a+qb_6}\ell b^0{_2}\right]b^2{_2}\,.
\eea
\esubeq
The existence of $\tG$ is in accordance with the above requirement (c).
A direct comparison with formulas (6.1) and (6.2) in Ref. \cite{x9}
reveals again the correspondence \eq{6.2}.

The values of the improved generators of time translations and spatial
rotations are given by the corresponding surface terms, which define
the conserved charges of the system---the energy and the angular
momentum, respectively:
\be
E=\int_0^{2\pi}d\vphi\cE^1\, ,\qquad M=\int_0^{2\pi}d\vphi \cM^1\, .
\ee
In particular, the energy and angular momentum of the BTZ-like black
hole \eq{4.1} are:
\be
E=\left(1+\frac{qb_6}{a}\right)m\, ,\qquad
M=\left(1+\frac{qb_6}{a}\right)J\, .
\ee
We have checked that Nester's covariant approach, see \cite{x23,x24},
yields the same result. The conserved charges depend on the curvature
strength $q$, but not on the torsion strength $p$. For $qb_6\ne 0$, the
values of the black hole charges differ from the corresponding GR
expressions.

\subsection{Asymptotic symmetry and black hole entropy}

Asymptotic AdS symmetry is described by the asymptotic canonical
algebra of the improved generators $\tilde G$. Expressed in terms of
the Fourier modes $L_n^\mp$ of $\tG$, this algebra has the form of two
independent Virasoro algebras:
\be
i\{L_n^\mp,L_m^\mp\}=
  (n-m)L_{n+m}^\mp+\frac{c^\mp}{12}n^3\d_{n,-m}\, ,
\ee
where $c^\mp$ are classical central charges:
\be
c^-=c^+=\left(1+\frac{qb_6}{a}\right)\frac{3\ell}{2G}\, .
\ee
Once we have the central charges, we can use Cardy's formula to
calculate the black hole entropy:
\be
S=\left(1+\frac{qb_6}{a}\right)\frac{2\pi r_+}{4G}\, ,
\ee
where $r_+$ is the radius of the outer black hole horizon.

With the above results for the conserved charges and entropy, one can
easily verify the validity of the first law of black hole
thermodynamics \cite{x10}.

\section{Concluding remarks}

In this paper, we studied dynamical structure of 3D gravity with
propagating torsion.

Starting with the quadratic parity-preserving PGT Lagrangian, we
derived the general form of the field equations and analyzed their
particle content. Then, focusing on the nonlinear aspects in the AdS
sector, we found the conditions on parameters for which the BTZ-like
black hole is an exact solution of the model. Finally, by examining the
AdS asymptotic structure, we obtained the expressions for the conserved
charges of the black hole with torsion, whereas the asymptotic symmetry
is found to be the conformal symmetry, described by two independent
Virasoro algebras with equal central charges.

In our next paper, we plan to explore the stability of the canonical
structure under linearization in order to identify viable versions of
the PGT model considered here.

\section*{Acknowledgements}

This work was supported by the Serbian Science Foundation under Grant
No. 171031.

\appendix
\section{Irreducible decomposition}
\setcounter{equation}{0}

The quadratic terms in the gravitational Lagrangian \eq{2.1} are
systematically defined with the help of the irreducible decomposition,
see \cite{x18,x6}.

The torsion has three irreducible pieces. In the basis of
orthonormal frames $b^i$, they take the form
\bsubeq
\bea &&{}^{(2)}T_i=\frac{1}{2}\eta_{ij}v_k b^jb^k\, ,      \nn\\
&&{}^{(3)}T_i=\frac{1}{2}a\ve_{ijk}b^jb^k\, ,              \nn\\
&&{}^{(1)}T_i=T_i-{}^{(2)}T_i-{}^{(3)}T_i
             =\frac{2}{3}t_{ijk}b^jb^k\, ,
\eea
where
\bea
&&v_i:=T^m{}_{mi}\, ,\qquad a:=\frac{1}{6}\ve_{mnr}T^{mnr}\,,\nn\\
&&t_{ijk}:=\frac{1}{2}(T_{ijk}+T_{jik} )
  +\frac{1}{4}(\eta_{ki}v_j+\eta_{kj}v_i)
  -\frac{1}{2}\eta_{ij}v_k\, ,
\eea
\esubeq
and $\ve^{ijk}t_{ijk}=t^i{}_{ik}=0$.

In $D=3$, the Weyl curvature vanishes, see \eq{2.10}, so that the RC
curvature $R_{ij}$ can be expressed in terms of the Ricci 1-form
$\hR_i=\hR_{ij}b^j$. As a consequence, $R_{ij}$ does not have six, but
only three irreducible parts:
\bsubeq
\bea
&&^{(4)}R_{ij}=b_i\sS_j-b_j\sS_i\, , \nn\\
&&^{(5)}R_{ij}=b_iA_j-b_jA_i\, , \nn\\
&&^{(6)}R_{ij}=R_{ij}-{}^{(4)}R_{ij}-{}^{(5)}R_{ij}
              =\frac{1}{6}Rb_ib_j \, ,
\eea
where $\sS_i$ and $A_i$ are irreducible parts of the Ricci 1-form:
\bea
&&\sS_i=\left(\hR_{(ij)}-\frac{1}{3}\eta_{ij}R\right)b^j\, , \nn\\
&&A_i=\hR_{[ij]}b^j\, .
\eea
\esubeq

\section{On the first-order formulation}
\setcounter{equation}{0}

In the first-order formulation \eq{5.1}, the form of the potential is
chosen so that when the field momenta $\t_i$ and $\r_i$ take their on
shell values, we return to the original formulation \eq{2.1}.

Let us first consider the form of the potential for the torsion squared
piece of the gravitational Lagrangian.  We start with
\bsubeq\lab{B.1}
\be
L_{T^2}=T^i\t_i-V_{\t^2}(b^i,\t^i)\, ,
\ee
where $V_{\t^2}$ is a 3-form quadratic in $\t_i$, conveniently written
as:
\be
V_{\t^2}=\frac{1}{2}\t^i\left[\m_1{}^{(1)}(\hd\t_i)
    +\m_2{}^{(2)}(\hd\t_i)+\m_3{}^{(3)}(\hd\t_i)\right]\,,\lab{B.1b}
\ee
\esubeq
where $\m_k$ are the parameters that depend on the torsion parameters
$a_k$ in \eq{2.1}. In the generic case when all $\m_k$'s are
nonvanishing, the variation of \eq{B.1} with respect to $\t_i$ yields
\be
T_i=\m_1{}^{(1)}(\hd\t_i)+\m_2{}^{(2)}(\hd\t_i)
    +\m_3{}^{(3)}(\hd\t_i)\, .                             \lab{B.2}
\ee
As a consequence, the Lagrangian \eq{B.1} takes the form
$L_{T^2}=\frac{1}{2}T^i\t_i$. Now, to show that this $L_{T^2}$ reduces
to its original form \eq{2.1b}, we have to solve Eq. \eq{B.2} for
$\t_i$. First, we note that \eq{B.2} can be broken into three
irreducible pieces: ${}^{(k)}T_i=\m_k{}^{(k)}(\hd\t_i),\, k=1,2,3$.
Then, the choice $\m_k=1/2a_k$ leads to
$^{(k)}(\hd\t_i)=2a_k{}^{(k)}T_i$, which implies:
\bea
&&\t_i=2\hd(a_1{}^{(1)}T_i+a_2{}^{(2)}T_i+a_3{}^{(3)}T_i)=H_i\,,\nn\\
&&L_{T^2}=\frac{1}{2}T^iH_i\, ,
\eea
in accordance with \eq{2.1}. On the other hand, if any $a_k$ vanishes,
the corresponding $\m_k$ must also vanish.

An analogous construction holds in the curvature sector, too.

\section{Reduced phase space}
\setcounter{equation}{0}

Starting from the basic PBs
$\{b^i{_\m},\pi_j{^\n}\}=\d^i_j\d^\n_\m\d(\mbox{\boldmath{$x-x'$}})$
etc., we find the nontrivial piece of the PB algebra between the
primary constraints $X_A:=(\phi_i{^\a},\Phi_i{^\a},p_i{^\a},P_i{^\a})$:
\bea
&&\{\phi_i{^\a},p_j{^\b}\}=-\ve^{0\a\b}\eta_{ij}\d\, ,\qquad
  \{\Phi_i{}^\a,P_j{}^\b\}= -\ve^{0\a\b}\eta_{ij}\d\, .
\eea
Now, we define the reduced phase space $R$ by eliminating the
momentum variables from the second class constraints $X_A$. To
construct the corresponding DBs, we consider the $24\times 24$ matrix
$\D$ with matrix elements $\D_{AB}=\{X_A.X_B\}$:
\be
\D(\vec x,\vec y)=
     \left( \ba{cccc}
                0& 0   & -1  &0   \\
              0& 0&  0 &-1\\
               -1&  0    &  0& 0\\
               0&-1&0&0
               \ea
     \right)\otimes\ve^{0\a\b}\eta_{ij}\d(\vec x-\vec y)\, .\nn
\ee
The matrix $\D$ is regular, and its inverse has the form
\be
\D^{-1}(\vec x,\vec y)=
     \left( \ba{cccc}
                       0 & 0 &1& 0 \\
                       0 &0& 0&1    \\
                     1& 0 &0&0  \\
                     0 & 1 &0&0
            \ea
     \right)\otimes\ve_{0\b\g}\eta^{jk}\d (\vec y-\vec z)\,.\nn
\ee
The matrix $\D^{-1}$ defines the DBs in the phase space $R$:
$$
\{\phi,\psi\}^*=\{\phi,\psi\}
                  -\{\phi,X_A\}(\D^{-1})^{AB}\{X_B,\psi\}\, .
$$
Explicit form of the nontrivial DBs is given in \eq{5.5}.

\section{Poincar\'e gauge transformations}
\setcounter{equation}{0}

In this appendix, we wish to show that the canonical generator $G$,
given by Eq. \eq{5.7}, produces the standard Poincar\'e gauge
transformations on the whole phase space $R$. To do that, we note that
the action of $G$ on the phase-space variables has the form
$$
\d_0\phi=\{\phi,G\}^*=-\{\phi,G_1\}^*-\{\phi,G_2\}^*\,.
$$

Starting with $b^i{_0}$ and $\om^i{_0}$, we find
\bea
&&\d_0 b^i{_0}=-\ve^{ijk}b_{j0}\th_k-\dot\xi^\mu b^i{_\mu}
  -\xi^\mu\pd_\mu b^i{_0}\,,                               \nn\\
&&\d_0 \om^i{_0}=-\nab_0\th^i-\dot\xi^\mu \om^i{_\mu}
  -\xi^\mu\pd_\mu \om^i{_0}\,,
\eea
as expected. For the field $b^i{_\a}$, we have:
\bea
&&\{b^i{_\a},G_2\}^*=\ve_{ijk}b^j{_\a}\th^k\, ,            \nn\\
&&\{b^i{_\a},G_1\}^*=(\pd_\a\xi^\mu)b^i{_\mu}
  +\xi^\mu\pd_\mu b^i{_\a}+\xi^\m X^i{}_{\a\m}\, ,         \nn\\
&&X^i{}_{\a\m}:= T^i{}_{\a\mu}
  +\ve_{0\a\b}\left(\t^j{_\m}\frac{\pd^2\cV}{\pd\t^j{_0}\pd\t_{i\b}}
  +b^j{_\m}\frac{\pd^2\cV}{\pd b^j{_0}\pd\t_{i\b}}\right)\,.\nn
\eea
By noting that the piece of $\cV$ that is quadratic in
$\t_{i\m}$ can be written in the form
$$
\cV_{\t^2}=\frac{1}{2}b\left(\m_1\t^{ij}\t_{ij}
  +\m_2\t^{ij}\t_{ji}+\m_3\t^2\right)\, ,
$$
see \eq{B.1b}, we find that $X^i{}_{\a\m}$ coincides with the field
equation \eq{5.2b}$_1$:
$$
X^i{}_{\a\m}=T^i{}_{\a\m}
  -\ve_{\a\m\n}\frac{\pd\cV_{\t^2}}{\pd\t_{i\n}}\, .
$$
Thus, on shell, we obtain the standard PGT transformation law:
\be
\d_0 b^i{_\a}\approx-\ve_{ijk}b^j{_\a}\th^k
  -(\pd_\a\xi^\mu)b^i{_\mu}-\xi^\mu\pd_\mu b^i{_\a}\, .
\ee
Application of the same procedure to $\om^i{_\a}$ yields:
\bea
&&\{\om^i{_\a},G_2\}^*=\nab_\a\th^i\, ,                    \nn\\
&&\{\om^i{_\a},G_1\}^*=(\pd_\a\xi^\mu)\om^i{_\mu}
  +\xi^\mu\pd_\mu \om^i{_\a} +\xi^\m Y^i{}_{\a\m}\, ,      \nn\\
&&Y^i{}_{\a\m}:= R^i{}_{\a\mu}
  +\ve_{0\a\b}\left(\r^j{_\m}\frac{\pd^2\cV}{\pd\r^j{_0}\pd\r_{i\b}}
  +b^j{_\m}\frac{\pd^2\cV}{\pd b^j{_0}\pd\r_{i\b}}\right)\,,\nn
\eea
whereupon \eq{5.2b}$_2$ leads to:
\be
\d_0 \om^i{_\a}\approx-\nab_\a\th^i
-(\pd_\a\xi^\mu)\om^i{_\mu}-\xi^\mu\pd_\mu \om^i{_\a}\,.
\ee

Next, we turn our attention to the covariant field momenta
($\t^i{_\m},\r^i{_\m}$). The calculations of $\d_0\t^i{_0}$ and
$\d_0\r^i{_0}$ are straightforward. To calculate $\d_0\t^i{_\a}$, we
start with:
\bea
&&\{\t^i{_\a},G_1\}^*=-(\pd_\a\xi^\m)\t^i{_\m}
  -\xi^\m\left(\nabla_\a\t^i{_\m}-\nabla_\m\t^i{_\a}
              +\ve_{0\a\b}f_\m{}^{i\b}\right)\, ,            \nn\\
&&f_\m{}^{i\b}:=b^j{_\m}\frac{\pd^2\cV}{\pd b^j{_0}\pd b_{i\b}}
  +\t^j{_\mu}\frac{\pd^2 \cV}{\pd \t^j{_0}\pd b_{i\b}}
  +\r^j{_\m}\frac{\pd^2 \cV}{\pd \r^j{_0}\pd b_{i\b}}\, .    \nn
\eea
Then, using the identity
$$
\ve_{0\a\b}f_\m{}^{i\b}=-\ve_{\a\m\n}\frac{\pd\cV}{\pd b_{i\n}}\, ,
$$
and the field equation \eq{5.2a}$_1$, we obtain the PGT transformation
law:
\be
\d_0\t^i{_\a}\approx-\ve^{ijk}\t_{j\a}\th_k
  -(\pd_\a\xi^\m)\t^i{_\m}-\xi^\m\pd_\m\t^i{_\a}\, .
\ee
Similar considerations lead to an analogous formula for
$\d_0\r^i{_\a}$.

Finally, to complete our considerations, we have to show that the
canonical generator $G$ produces the correct local Poincar\'e
transformations of the momentum variables in $R$. These transformations
are determined by the defining relations $\pi_A=\pd\tcL/\pd\dot\vphi^A$
and the known transformation laws for $\vphi^A$ and $\tcL$, as
described in Ref. \cite{x29}. For $\pi_i{^0}$, for instance, this
method yields
\bsubeq
\bea
\d_{\rm P}\pi_i{^0}&=&-\ve_{ijk}\pi^{j0}\th^k
  +(\pd_0\xi^0-\pd_\a\xi^\a)\pi_i{^0}-\xi^\r\pd_\r\pi_i{^0} \nn\\
  &&+(\pd_\a\xi^0)( \pi_i{^\a}+H_i{}^{\a 0})\, ,
\eea
where $H_i{}^{\a 0}=-H_i{}^{0\a}=-\ve^{0\a\b}\t_{i\b}$. Note that the
last term, which is proportional to the primary constraint
$\phi_i{^\a}$ in \eq{5.3}, vanishes identically on $R$. Now, we wish to
compare this result with the canonical action of $G$ on $\pi_i{^0}$,
$\d_0\pi_i{^0}=\{\pi_i{^0},G\}^*$. Explicit calculation yields
\bea
&&\d_0\pi_i{^0}=\d_{\rm P}\pi_i{^0}+\xi^\m f_{i\m}\, ,        \\
&&f_{i\m}:=b^j{_\m}\frac{\pd^2\cV}{\pd b^j{_0}\pd b^i{_0}}
  +\t^j{_\m}\frac{\pd^2\cV}{\pd\t^j{_0}\pd b^i{_0}}
  +\r^j{_\m}\frac{\pd^2\cV}{\pd\r^j{_0}\pd b^i{_0}}\, .\nn
\eea
\esubeq
To obtain this result, we used the relations
\bea
&&\{\pi_i{^0},\pd_0 b^j{_0}\}^*=
  \{\pi_i{^0},\{b^j{_0},H_T\}^*\}^*=0\, ,                  \nn\\
&&\{\pi_i{^0},\hcH_j\}^*=
          \frac{\pd^2\cV}{\pd b^j{_0}\pd b^i{_0}}\, ,      \nn
\eea
and similarly for $\{\pi_i{^0},\hat\cT_j\}^*$ and
$\{\pi_i{^0},\hat\cR_j\}^*$. For the generic form of $\cV$, the term
$f_{i\m}$ vanishes, which leads to $\d_0\pi_i{^0}\approx \d_{\rm
P}\pi_i{^0}$, as expected. Showing that the analogous results hold for
$(\Pi_i{^0},p_i{^0},P_i{^0})$, we complete the proof that the gauge
generator \eq{5.7} acts correctly on the whole phase space $R$.


\end{document}